\documentclass[10pt]{IEEEtran}
\ifCLASSINFOpdf

\else

\fi
\usepackage{amsfonts}
\usepackage{array}
\usepackage{amssymb}
\usepackage{amsmath}
\usepackage{graphicx}
\usepackage{amsfonts}
\usepackage{array}
\usepackage{color}
\usepackage[]{cite}
\usepackage{amssymb}
\usepackage{tabularx}
\usepackage{mathrsfs}
\usepackage{amsthm}
\usepackage{multirow}
\usepackage{bpchem}
\usepackage{subfigure}
\usepackage{float}
\usepackage{caption}
\usepackage[]{cite}
\begin{document}
\pagestyle{empty}
%
\title{IEEE 802.15.6-based Prototype System for WBAN: Design and Implementation}
%
\author{\IEEEauthorblockN{Xiaonan Su, Changle Li*, Xiaoming Yuan}\\
\IEEEauthorblockA{State Key Laboratory of Integrated Services Networks, Xidian University, Xi'an, Shaanxi, 710071 China\\
*clli@mail.xidian.edu.cn}
}

\maketitle
\thispagestyle{empty}

\begin{abstract}
Various current trends such as ever growing population and accelerated aging effects have effectively promoted the growth of Wireless Body Area Network (WBAN). Being the specialized standard of WBAN, the burgeoning IEEE 802.15.6 is on its way to refinement and perfection. Since experiments are the foundation of research in WBAN that reproduce system operating patterns, yet existing platforms in WBAN are mainly simulations that could not highly conform to reality with strict requirements to some degree. Thus, a reliable and efficient system based on the IEEE 802.15.6, named WBAN prototype system, is proposed and implemented in this paper. The prototype system is ascertained to be valid, authentic and efficient via analyses of scenario tests. Running in circumstances much closer to reality, the system obtains experimental results that meet rigid requirements preferably. Furthermore, based on the valid platform we created, optimizations of two parameters in the IEEE 802.15.6 by actual measurements are derived. We obtain an optimum number of transmission retries and relationship of payload length within 30 with FER, making a trade-off between waste of resource and high packet loss rate.
\end{abstract}


%
\IEEEpeerreviewmaketitle

\section{Introduction}
Recent years have witnessed the rapid growth of global population and aggravated aging effects\cite{javaid2013ubiquitous}, which results in urgent demand of comprehensive health care systems, yet medical resources cannot always satisfy our needs. In addition, a long-term and comprehensive health care is usually so costly that normal people cannot afford it. Furthermore, even in sub-health condition, people are eager for more freedom and personalized intelligent lifestyle with medical equipment. Thus, providing an efficient, convenient and low-cost health care system, Wireless Body Area Network (WBAN) is gaining an increasing attention\cite{talha2015priority}\cite{movassaghi2014wireless}.

WBAN is a kind of body-centered short-range communication technology that connects and interacts with various sensors located inside, on the surface or outside of the human body, monitoring the human body conditions and the surrounding environments\cite{gu2013implementation}\cite{du2016beam}. It is an emerging cross-technology which has a close relationship with Wireless Personal Area Network (WPAN), Wireless Sensor Network (WSN)\cite{rajeshwari2015hierarchical} and sensor technology\cite{rawat2014wireless}.

Existing simulations of WBAN can be clarified into three categories: software, hardware, and their combination. Popular simulation software for networks includes Optimized Performance Network Engineering Tool (OPNET), Objective Modular Network Test-bed in C++ (OMNET++) and Network Simulator (NS). OPNET provides a skeleton of 14-stage radio pipeline. OMNET++ is a discrete event simulation environment primarily designed for communication networks\cite{bause2010simulation}. Network Simulator Version 3 (NS-3)\cite{mastorakis2015ndnsim} has been applied to simulating systems in a highly controlled environment of WBAN. However, there are some discrepancies between simulations on software and actual results in realistic environments with time-variant, fading and susceptibility to disturbance characteristics. Some propose a cooperative software-hardware approach for WBAN implementation\cite{chen2014cooperative} that decreases the time and complexity of implementation. However, it is still a kind of makeshift which will be pale in reality and correctness before complete hardware platforms to some degree, since it is hardware that applications and protocols finally run on. Hardware platforms can propose models with comprehensive evaluations and protocols superior in performance and practical in application\cite{ge2015energy}. Nevertheless, to our best knowledge after serious investigation, few WBAN or IEEE 802.15.6 module has been constructed on hardware in public.

 Due to this issue, this paper realizes the WBAN prototype system based on IEEE 802.15.6, and makes research on its actual performance. Specifically, contributions of this paper are presented as follows:
\begin{itemize}
\item
We not only design and implement the WBAN modules such as PHY module, MAC module, application module and security module, but also assemble them via primitives we defined to work coordinately as a prototype system based on IEEE 802.15.6.
\item
We carry out several sets of scenario tests under different parameters to verify the validity of our prototype system.
\item
We analyze testing results to optimize two parameters in IEEE 802.15.6. We have derived the optimal maximum transmission retries and payload length within 30, moreover, the relationship between payload and FER is excogitated. These not only provide a vital reference to parameters in the standard IEEE 802.15.6, but also supplement to the standard in several specific scenes.
\end{itemize}

The rest of the paper is organized as follows. Section II briefly reviews the most relevant and basic contents in IEEE 802.15.6. Section III elaborates the scheme design of WBAN prototype system and its implementation. Section IV performs a set of scenario tests, makes an evaluation of the prototype system, and studies the optimal transmission retries and the variation of FER with payload. Finally, section V concludes the paper.

\section{An Overview of IEEE 802.15.6}
The IEEE 802.15.6 standard\cite{chang2015adaptive} is especially designed for WBAN by modifying the PHY and MAC parameters similar to the IEEE 802.15.4 standard\cite{hamalainen2015etsi} to support short range, ultra-low power and reliable wireless communication in vicinity or inside of living tissue\cite{khan2015performance}. A WBAN consists of a sole hub and several nodes whose number ranges from 0 to 64. The nodes are used to collect some parameters of the body, while the hub executes special actions according to the data received or interactions with the user\cite{kannan2008robust}.

According to the reference model of IEEE 802.15.6, the hub and nodes are divided into two layers: PHY layer and MAC layer, as depicted in Fig. 1. The service access point (SAP) is an interface between two layers. MAC layers provide service for application layers through MAC SAPs, while PHY layers provide service for MAC layer through PHY SAPs. When transmitting data, service data units exchange between PHY layer, MAC layer and application layer through respective SAPs.

\begin{figure}[!htb]
\centering
\includegraphics[width=0.48\textwidth]{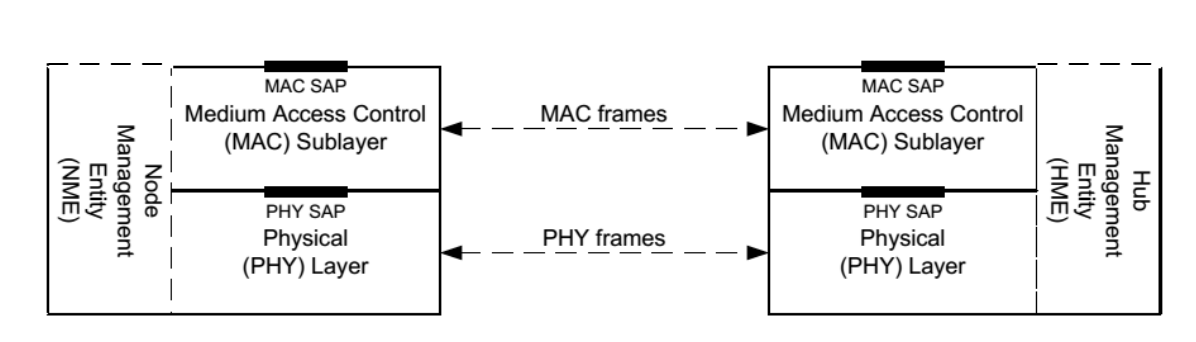}
\caption{Reference model in IEEE 802.15.6}
\label{fig_Reference model in IEEE 802.15.6}
\end{figure}

\section{Scheme Design and Implementation}
This section focuses on the ground-breaking exploration of a WBAN prototype system. System architecture and full implementation are to be specified next.
\subsection{System Architecture}
The whole WBAN prototype system is complicated in both structure and function, whose content includes realizations of the application layer, MAC layer, physical (PHY) layer and security certification.

 We design and construct the hub and nodes according to the communication mechanism of WBAN. A node consists of a sensor board, DSP, Field Programmable Gate Array (FPGA), a Radio Frequency (RF) board and an antenna, while a hub is composed of a Bluetooth board, DSP, FPGA, a RF board and an antenna, as illustrated in Fig. 2. Both of them contain the application layer, MAC layer, physical layer as well as security certification section.

\begin{figure}[!htb]
\centering
\includegraphics[width=0.45\textwidth]{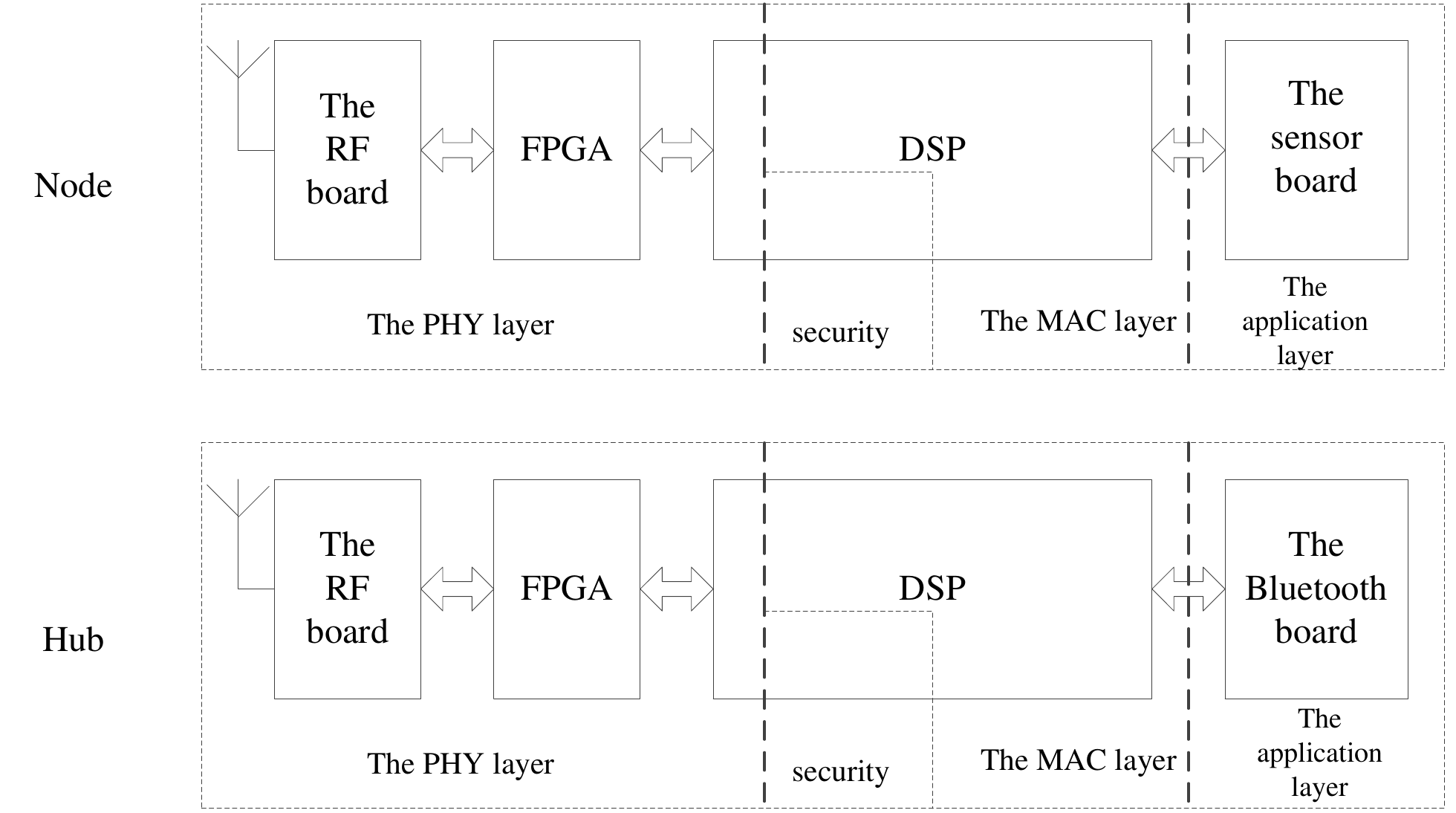}
\caption{ Hardware structure of a node and a hub}
\label{fig_Hardware structure of a node and a hub}
\end{figure}




DSP accomplishes a series of data control and buffering between MAC and PHY layer through FPGA. In addition, by cooperation with FPGA, hardware system with complex interactions could be achieved. The RF board with antenna is in charge of communication between the hub and nodes. The Bluetooth board in hub takes charge of transmitting the data collected to corresponding terminals via Bluetooth technique.

The implementation of WBAN prototype system is operated by polling program that consists of three modules: management service module, data service module and data transmission module, all of which are mainly achieved on DSP. Polling procedure is run as follows. After being charged, the system enters initialization part. Then, it enters the polling section that senses whether data from upper layers arrives. If it does arrive, different interfaces arrival events will trigger corresponding disposal mechanisms, otherwise, states and state transition events in the MAC layer will be polled and disposed respectively. Interface modules send primitives to management service module, data service module and data transmission module.

\subsection{Implementation}
Since the MAC layer locates in the functional centre of the system architecture and interacts with every other part in the system, we give it more weight. It is developed on Digital Signal Processor (DSP) and can be divided into three parts: scheme design, software realization and hardware implementation. The accomplished hub is shown in Fig. 3, which has much similarity with nodes. The platform provides ample interfaces to access various sensors, such as body temperature transducer and heart rate sensor that have already been used.

\begin{figure}[!htb]
\centering
\includegraphics[width=0.30\textwidth]{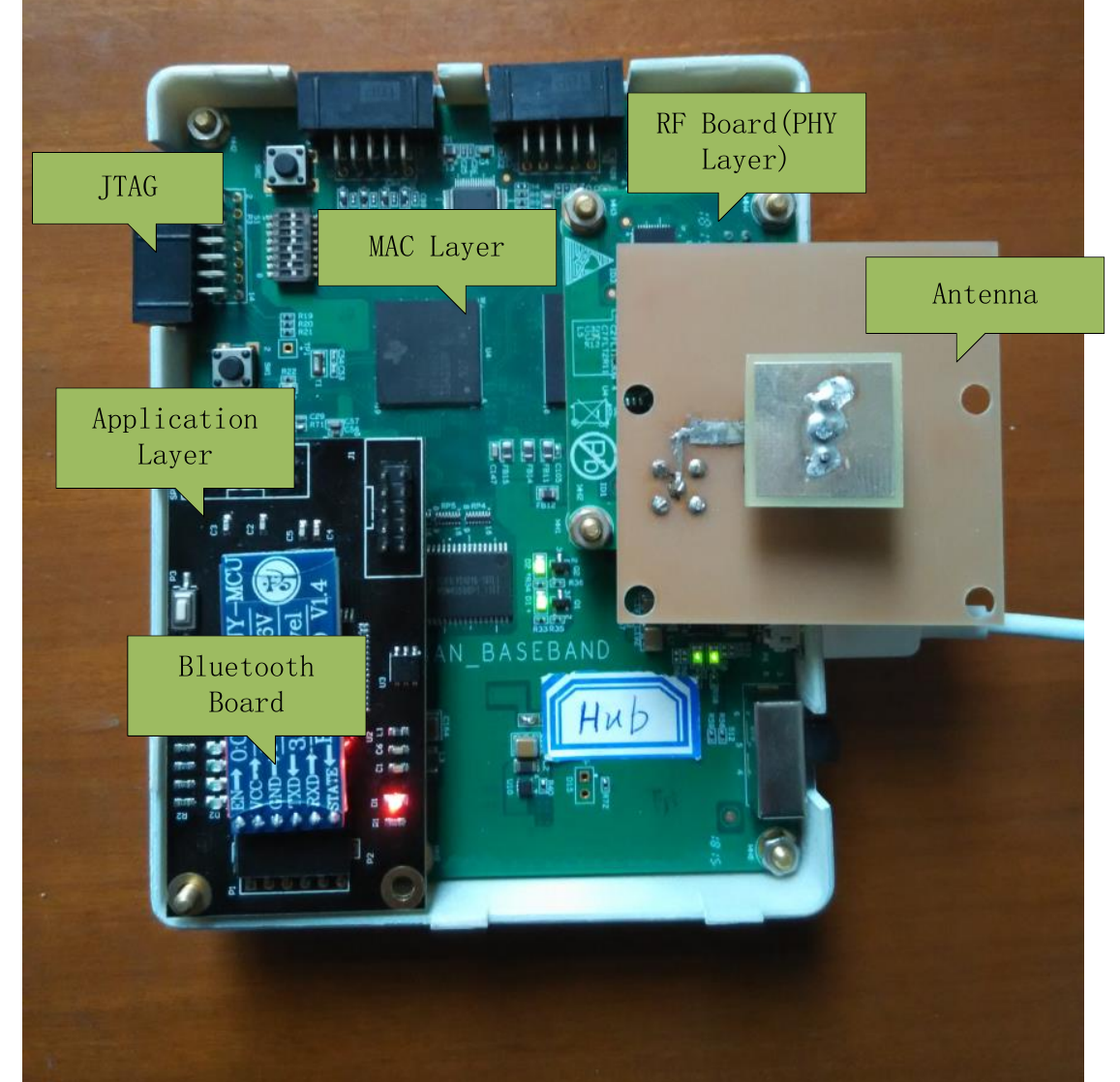}
\caption{Hub}
\label{Hub}
\end{figure}

All modules or layers in our prototype system communicate with each other using primitives. Therefore, we set about discussing primitives and module implementation.

\subsubsection{Primitives}
In order to transmit commands quickly and efficiently between modules and fill in the blank of explicit primitives specifications in IEEE 802.15.6, we design a whole set of interface primitives. The primitives are classified into three parts: management primitives, data service primitives and data transfer primitives, corresponding to the three modules in polling program which are to be discussed next.

\subsubsection{Management Module}
The management module takes charge of the connection establishment for networking, such as initialization, connection building and breaking. After receiving and processing management primitives, management service module constructs various types of management frames and sends them to the data transmission module. Meanwhile, it handles multiple types of management frames and constructs different management primitives sent to the application layer.


\subsubsection{Data Service Module}
Data service module is in charge of processing local nodes data such as data fragmentation, reconstitution, framing, etc. It constitutes data frames by receiving and disposing data service primitives. In addition, it receives and processes data frames, constructs the data service primitives, and sends them to finish the transmission of business data.

\subsubsection{Data Transmission Module}
Data transmission module is responsible for sending and receiving data frames, which covers encryption and decryption, CRC check, address judgment, access control, etc. It writes the CRC check in and encrypts the data frames, then sends them to the physical layer. Meanwhile, it receives and processes data, and uploads it after data transmission.

\subsection{Theoretical Analysis}
FER and PER are selected to examine the performance of the WBAN prototype system\cite{ta2009giant}. The FER is the ratio of error frames in transmission, and PER is the ratio of error packets in the transmission of sending packets which includes packet loss rate. High FER and PER indicate weak adaptability of system. If so, we should improve the system by adjusting relevant parameters.

The FER can be calculated by formula (1), where $S_{frm}$ and $R_{frm}$ represent the numbers of frames sent in nodes and frames received in the hub respectively. The PER can be figured out by formula (2), where $S_{pkt}$ and $R_{pkt}$ signify the numbers of packets sent and received respectively.

\begin{align}
P_{FER}&=(S_{frm}-R_{frm})/S_{frm}\\
P_{PER}&=(S_{pkt}-R_{pkt})/S_{pkt}
\end{align}

\subsubsection{Transmission retries}
Transmission retries is quite an important factor in wireless transmission. Setting it too high leads to much waste time and energy consumption, while too low may result in communication outage\cite{mao2006wsn06}. However, the IEEE 802.15.6 standard has not specified how to set the transmission retries. Therefore, we have studied the relationship between PER and transmission retries. On account of the length difference between data frames and acknowledgement (Ack) frames, we cannot neglect error Ack frames in performance investigation. In natural randomized environments, we assume that the PERs of data frames and Ack frames are the same. If a data frame is sent successfully over $m$ times transmission, its maximum transmission retries is $(m-1)$. The possibility of a successful transmission with $(m-1)$ transmission retries can be calculated by the following equations.

\begin{gather}
P_{suc}=C*F_{suc}\\
F_{suc}=(1-P_{fer})^2\\
C=\sum^{m-1}_{i=0}C_{m-1}^i[P_{fer}*(1-P_{fer})]^{m-i-1}*P_{fer}^i
\end{gather}
$F_{suc}$ represents the possibility of successful transmission of data frames and Ack frames, $C$ means the combinational number of successful transmission, and $P_{fer}$ signifies FER. The relationship diagram is present in Fig. 4. Enlarging and having a close observation of the figure, we can derive at least two conclusions. First, PER reaches its minimum value when $m$ ranges from 4 to 30. In addition, the possibility of successful transmission decreases with $m$. Thus, we can draw the conclusion that 4 is the appropriate value of $m$ for a minimum PER, namely, 3 is the proper maximum transmission retries. This conclusion will be certificated by our tests in the following section.
\begin{figure}[!htb]
\centering
\includegraphics[width=0.45\textwidth]{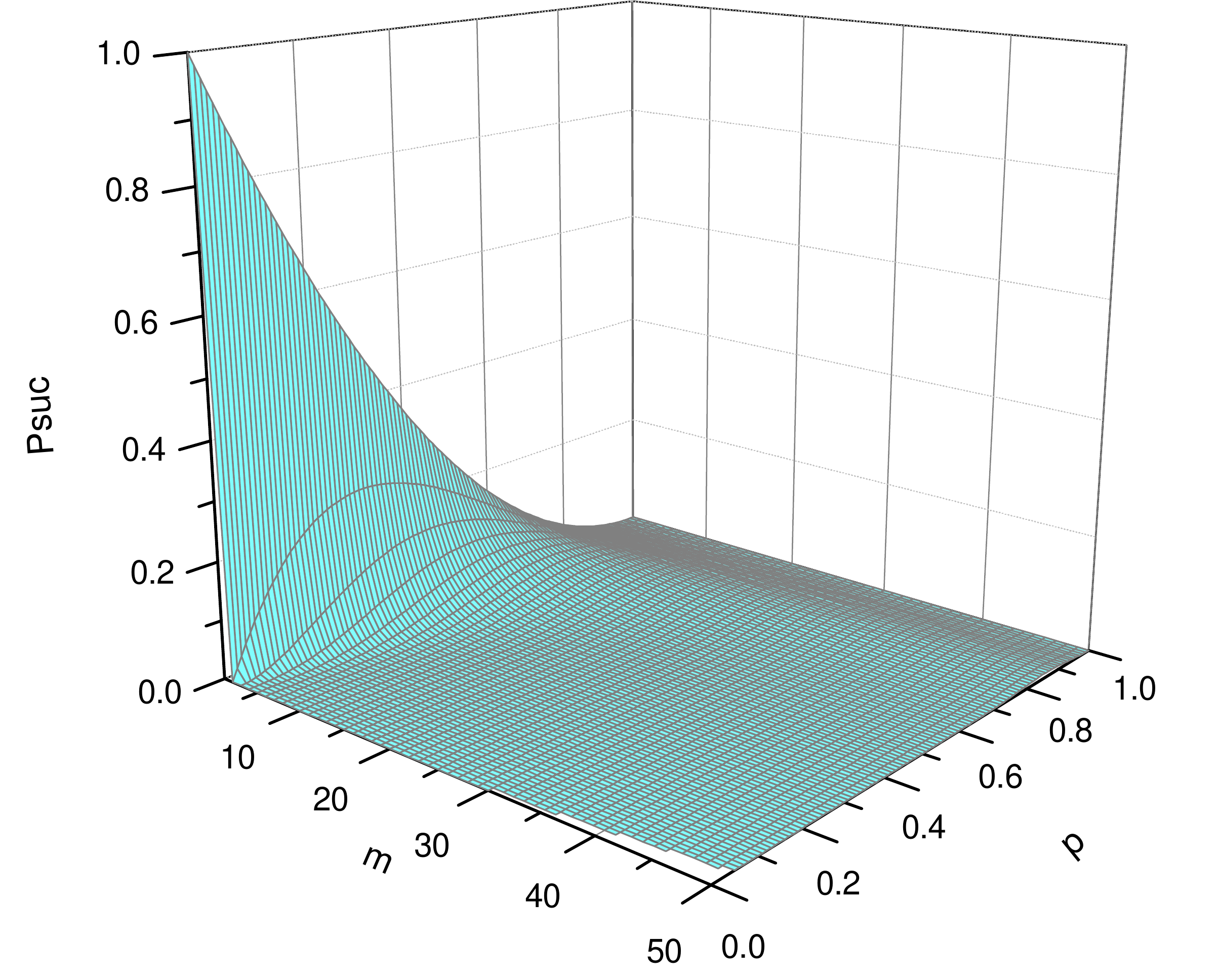}
\caption{Relationships between successful transmission possibility, maximum transmission retries and PER}
\label{fig_Relationships between successful transmission possibility, maximum transmission retries and PER}
\end{figure}

\subsubsection{Payload}
Payload, as a significant element in data transmission, is another parameter that we focus on. If it is set too low, the value of payload will lead to inefficiency and time-consumption\cite{mao2009graph}. While setting too high results in higher packet loss rate, since larger packets keep the channel in busy condition longer \cite{cavallari2014survey}. Thus, how to select a appropriate size of payload in WBAN becomes a non-ignorable problem.

In our prototype, there are eight bytes apart from payload in a data frame, and the length of an Ack frame is nine bytes. Since the length of a data frame is comparable to that of an Ack frame, the length of Ack frames cannot be neglected. Suppose that $P_{ber}$, $L_{ack}$ and $j$ represent bit error rate (BER), length of Ack frames and transmission times respectively, and $P_{ber}$ is a fixed value. Evidently, $L_{ack}$ obeys the geometric distribution and we can work the Eq. (6) to (8) out. The variation of $payload$ with $P_{ber}$ and $FER$ is illustrated in Fig. 5. We can learn from the diagram that $FER$ gets larger with the increase of $payload$ on a given $P_{ber}$. Then we exploit interior point algorithm to search an optimal evaluation of FER, as shown in the Eq. (9) to (13). Corresponding to the Fig. 5, when payload varies from 0 to 30, the payload declines with the decrease of FER. This verdict is also certificated in later tests.

\begin{gather}
L_{Ack}=8*9*j*P_{ber}^{j-1}(1-P_{ber}),\quad j=1,2,3,4,5\\
L_{data}=8*(payload+8)\\
FER=1-(1-P_{ber})^{L_{data}+L_{Ack}}
\end{gather}

\begin{figure}[!htbp]
\centering
\includegraphics[width=0.42\textwidth]{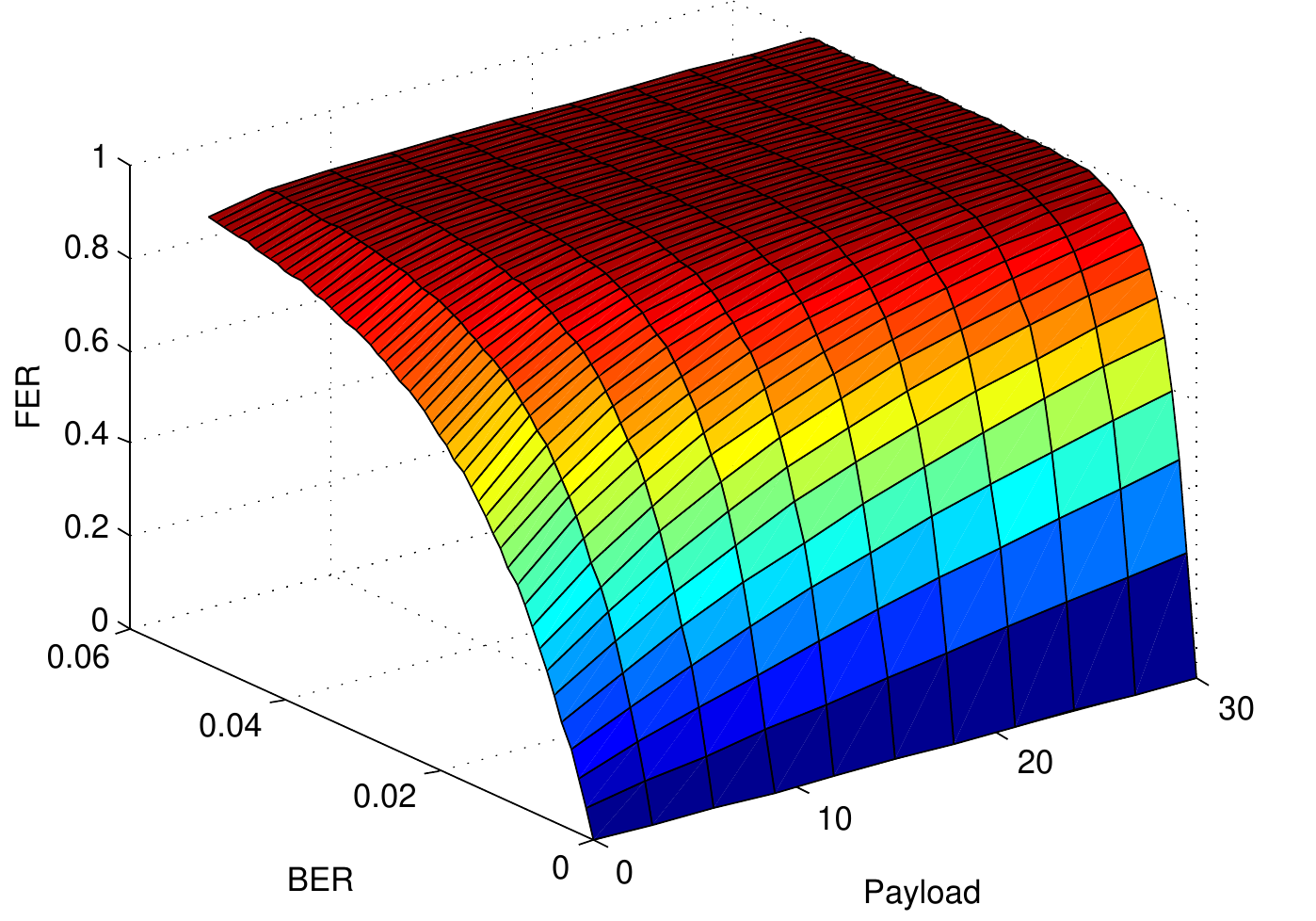}
\caption{Variations of FER with payload and BER}
\label{fig_Variations of FER with payload and BER}
\end{figure}
\begin{equation}
  \left\{ {\begin{array}{*{20}{l}}
{{\rm{min}}\;\;FER = 1 - {{(1 - {P_{ber}})}^{{L_{data}} + {L_{Ack}}}}}\\
{s.t.\;\;\;payload \geqslant 0}
\end{array}} \right.
\end{equation}

\begin{gather}
B(x)=\frac{1}{payload}\\
I(payload;\varepsilon_k)=(1-P_{ber})^{L_{data}+L_{Ack}}+\frac{\varepsilon_k}{payload}\\
I'(payload)=-8ln(1-P_{ber})(1-P_{ber})^{payload}-\frac{\varepsilon_k}{payload^2}\\
\varepsilon_k=-8payload^2*[ln(1-P_{ber})*(1-P_{ber})^{payload}]
\end{gather}

\section{Scenario Tests and Performance Evaluation}
In order to verify the validity and effectiveness of our system, scenario tests are executed. In this section, we first design the test scheme, then record experimental data. Influences of parameters are considered, such as the effect of distance on FER and the impact of retransmission on PER. Lastly, the performance of the WBAN prototype system is analyzed and some conclusions are drawn.
\subsection{Tests Design and Execution}

In the testing scheme, module testing is carried out first, then system testing follows. We employ six nodes and one hub to consist a typical communication system with star topology, trying our best to match with realistic applications owing to the constraints of hardware and floor space. Since WBAN scenarios are confined to within 10m and two-hop at most, among which seven nodes using single-hop within 5m just as our testing environment has certain representative significance. The nodes collect data from sensors, and the hub receives the data sent from the nodes. We can observe variables via the Code Composer Studio (CCS) software, an integrated development environment for DSP embedded software design.

\begin{table}[!t]
\renewcommand{\captionlabelfont}{\footnotesize}
\renewcommand{\arraystretch}{1.3}

    \centering
    \footnotesize
    \caption{\footnotesize  Parameters in Testing Scenario }

\begin{tabular}{|c|c|}
\hline
\textbf{Parameters} & \textbf{Value}\\
\hline
Frequency Band & 2.4 GHz\\
\hline
Average Time of One Test & 40 min\\
\hline
Data Transmission Rate & 121.4 kbps\\
\hline
Testing Distances & 1,2,5,10 m\\
\hline
Transmission retries & 0,1,2,3,4\\
\hline
Payload Length & 5,10,15,20,25,30 bytes\\
\hline
\end{tabular}
\end{table}

We employ distance as one variable, because degradation in network performance caused by the coexisting of BANs using same frequency band can be alleviated by distance. We confine the distance between node and hub to 5 meters, and designate the number of transmission retries from 0 to 4. Then we carry out a series of tests altering transmission distance and transmission retries respectively.
\subsection{Testing Results and Evaluation}
The analysis of system testing results include three steps. Firstly, process the experimental data into parameters such as FER and PER. Secondly, checkout whether the environment is configured properly. Finally, examine the recorded results.

%
The variation of FER with transmission distance is illustrated in Fig. 6. The FER stays low within the distance of 4m; FER is about 0.3\% in wired environment (connecting hub with nodes through wires, which is expected to be error-free as a reference) and about 0.5\% in wireless environment. Therefore, we can know that the transmission performance is fine within 4m and ordinary over 5m. It is obvious and reasonable that FER grows with distance. Since WBAN is a kind of short distance wireless networks, the results meet the requirements of the WBAN prototype system.

\begin{figure}[!htb]
\centering
\includegraphics[width=0.46\textwidth]{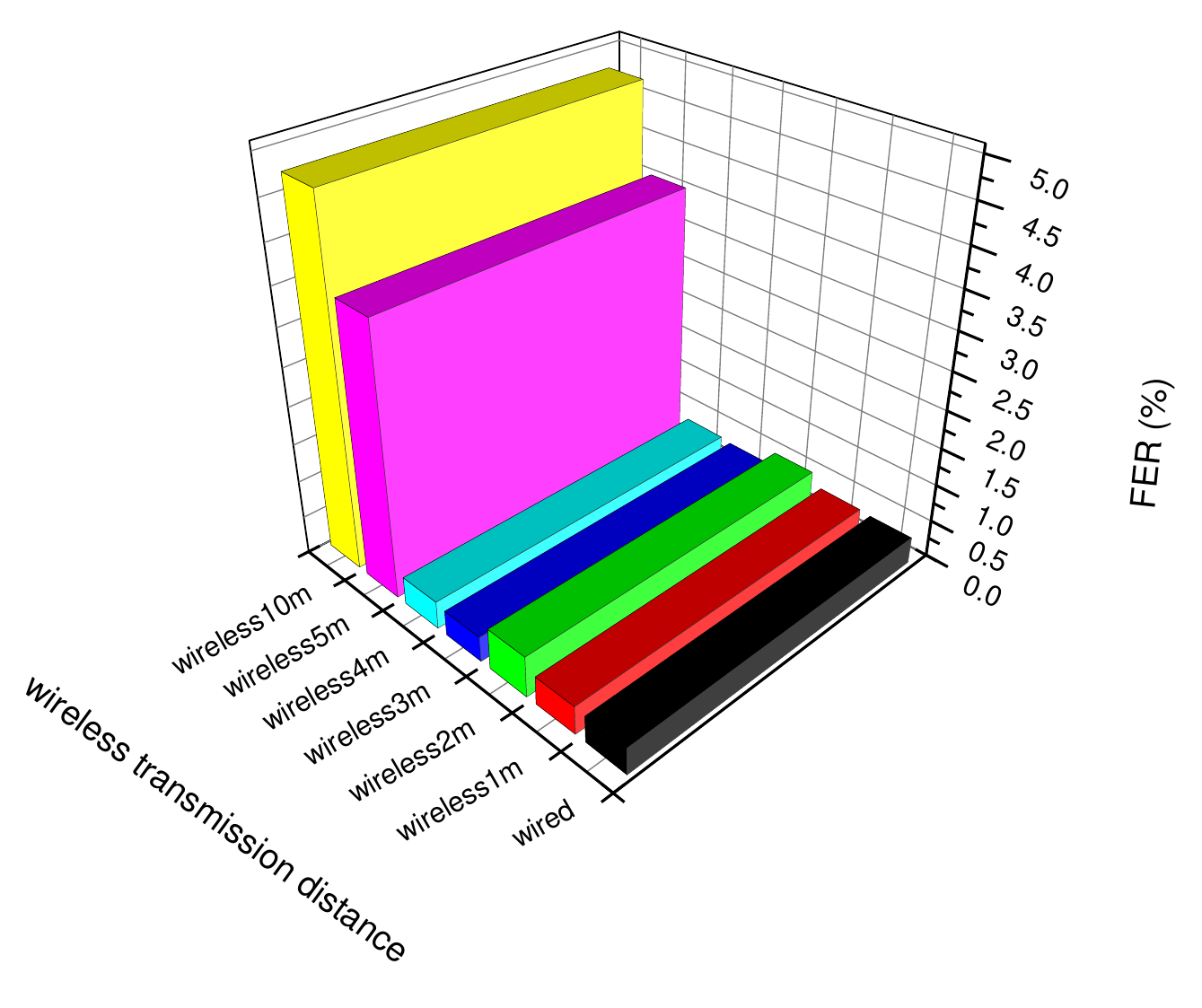}
\caption{Variations of FER with transmission distance}
\label{fig_Variations of FER with transmission distance}
\end{figure}

In order to validate the conclusion that the optimum number of transmission retries is 3, we conduct a series of experiments on our prototype system. We set five different numbers of transmission retries over four distances to observe variations of PER. The PER changes over transmission distances and transmission retries, as illustrated in Fig. 7, in which one point represents one experiment. We can learn from the diagram that the PER decreases with the maximum transmission retries increase. When maximum transmission retries is higher than 3, PERs of different distances are zero, which indicates no error packet. Therefore, 3 is the optimum maximum transmission retries for our prototype system, which verifies the theoretical analysis above.

\begin{figure}[!htb]
\centering
\includegraphics[width=0.36\textwidth]{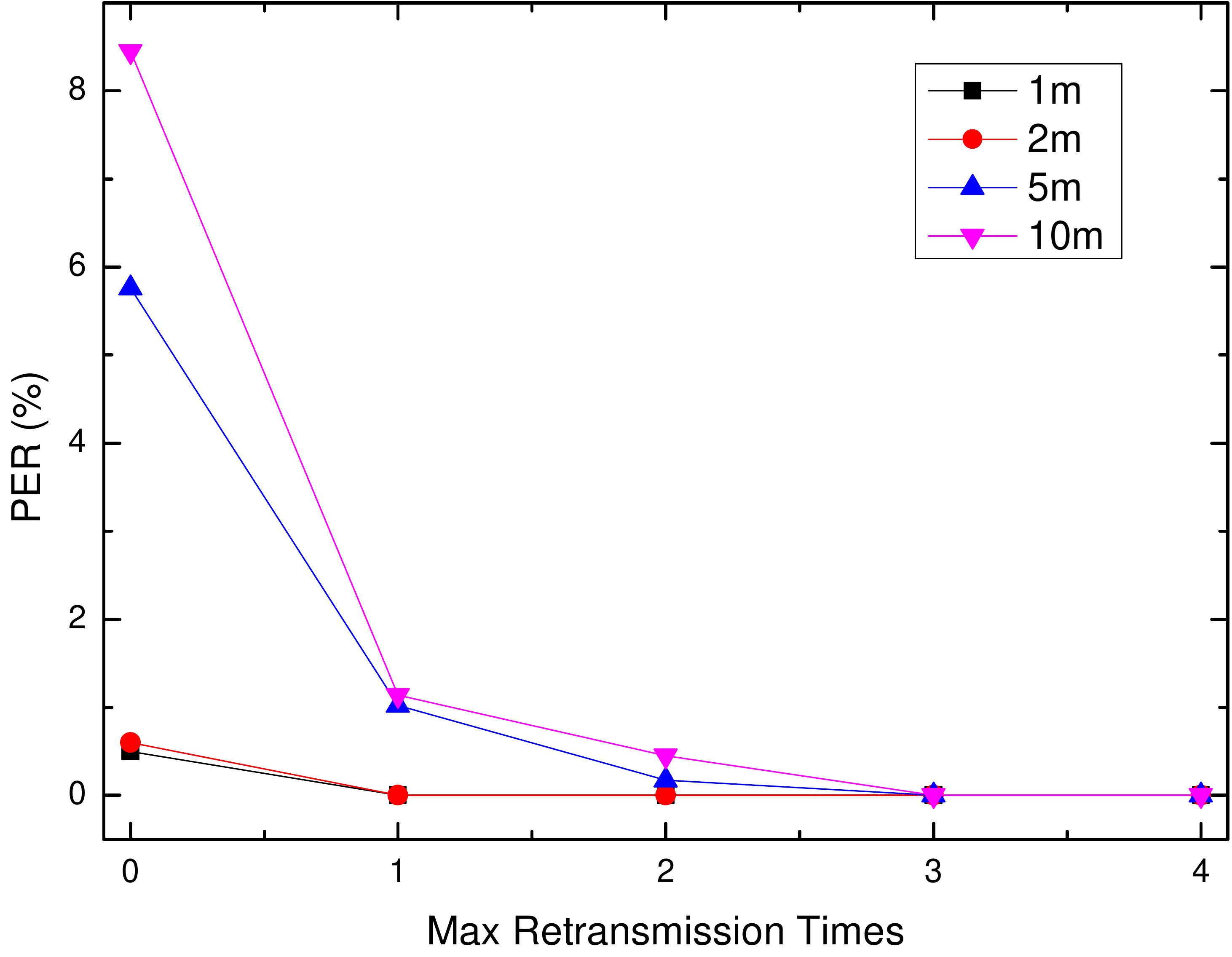}
\caption{Variations of PER with transmission distance and transmission retries}
\label{fig_Variations of PER with transmission distance and transmission retries}
\end{figure}
To verify the conclusion that payload declines with the decrease of FER when payload varies from 0 to 30, we conduct a series of tests in which FER varies with payload size and different distances commonly used in WBAN. As illustrated in Fig. 8, the larger number of retransmission retries is, the lower PER will be. FER is comparatively low in four different distances when payload length is around 10. FER increases with payload when other conditions are the same, namely, the accuracy of information will be less ideal with the payload increase in the content from 0 to 30. This diagram also coincides with Fig. 5. Owing to all unpredictable adverse factors in actual scenes, system performance in Fig. 8 is less preferable but more realistic. This is another aspect in optimizing the IEEE 802.15.6, and it will probably do much better if given a fuller development.
\begin{figure}[!htb]
\centering
\includegraphics[width=0.36\textwidth]{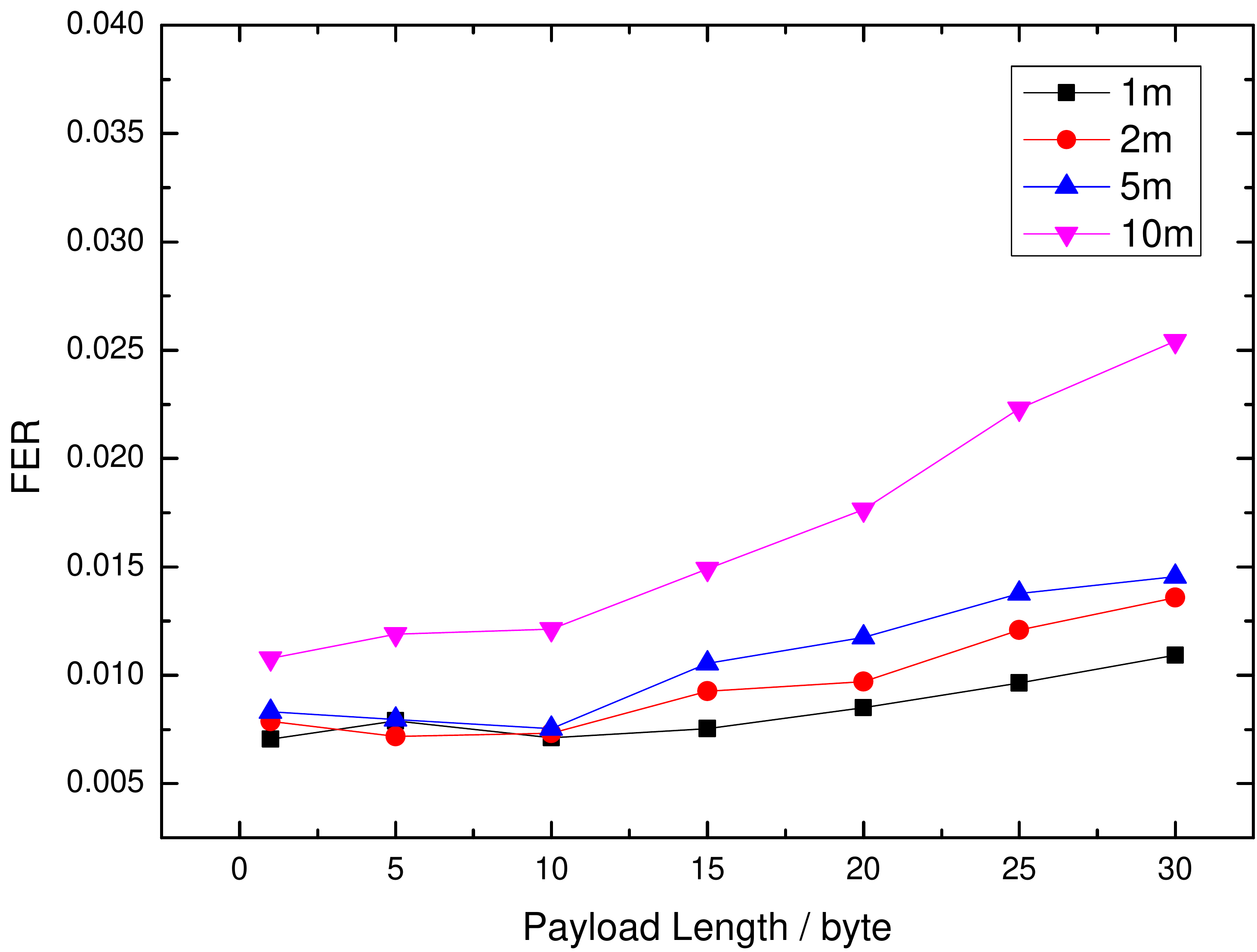}
\caption{Variations of FER with transmission distance and payload size}
\label{fig_Variations of FER with transmission distance and payload size}
\end{figure}
\section{Conclusions}
In order to provide a more realistic and effective WBAN platform as well as contribute to supplementing and optimizing the IEEE 802.15.6, we designed a WBAN prototype system and constructed the whole system from module to infrastructure. Then we created a set of interface service primitives for interaction between layers. In addition, we made analyses of its performance on the basis of scenario tests, and drew the conclusion that the system we designed can meet the demands of WBAN applications basically, and the optimal maximum number of transmission retries is 3, which ensures both FER and PER to be comparatively low. Furthermore, we worked out and validated the verdict that FER gets larger with payload within 30.





\bibliographystyle{IEEEtran}


%


%


\end{document}